\documentstyle[epsfig]{aipproc}
\begin{document}
\title{Methods of quasiclassical Green's functions in the theory of transport
phenomena in superconducting mesoscopic structures}
\author
 {
  A.F.Volkov$^*$ and V.V.Pavlovskii$^{\dagger}$
 }
\address
 {
  $^*$School of Physics and Chemistry, Lancaster University, Lancaster LA1 4YB and
   Institute of Radioengineering and Electronics of the Russian
   Academy of Sciencies, Mokhovaya str.11, Moscow
   103907, Russia\\
  $^{\dagger}$
  Institute of Physics and Technology of the Russian
  Academy of Sciences, Nakhimovskii Avenue 34, Moscow
  117218, Russia
 }

\maketitle

\begin{abstract}
A short introduction to the theory of matrix quasiclassical
Green's functions is given and possible applications of
this theory to transport properties of mesoscopic superconducting-
normal metal (S/N) structures are considered. We discuss a simplified
version of these equations in the diffusive regime and in the
case of a weak proximity effect. These equations are used for the
calculation of the conductance of different S/N structures and
for analysis of kinetic phenomena in these structures.
We discuss the subgap conductance measured
in SIN tunnel junctions and the mechanism of a nonmonotonic
dependence of the conductance of a N wire on temperature $T$
and voltage $V$, observed in an S/N structure.

Long-range, phase-coherent effects are studied in a 4-terminal
S/N/S structure under conditions when the Josephson critical
current is negligible (the distance between superconductors is
much larger then the coherence length in the normal wire). It
is shown that the Josephson effects may be observed in this system
if a current  $I$, in addition to a current $I_{1}$ in the S/N/S circuit,
flows through the N electrode.
\end{abstract}

\paragraph*{1.Introduction\protect\\}

\begin{figure}[b!] 
\centerline{\epsfig{file=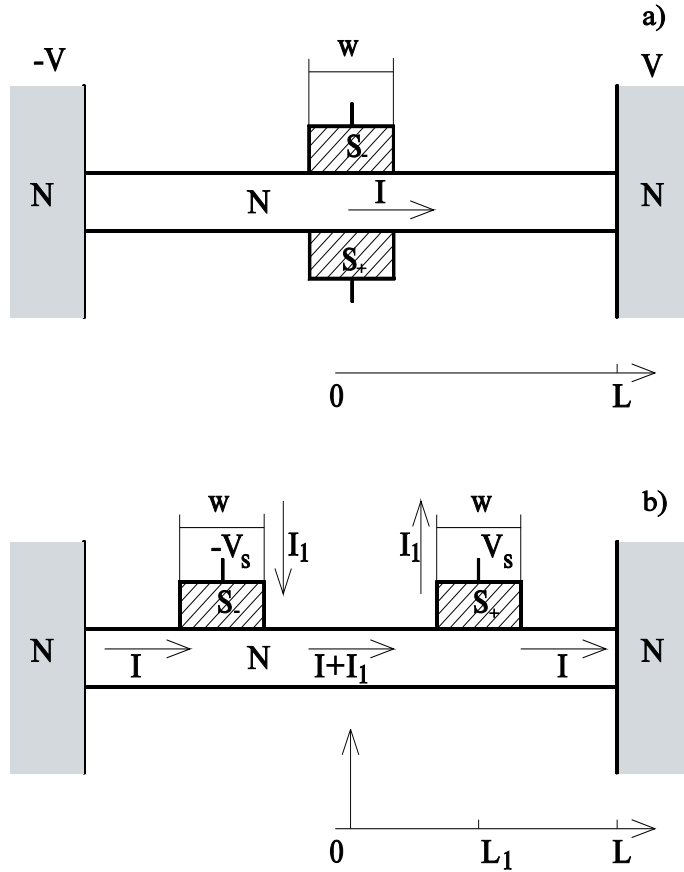,height=4.725in,width=3.5in}}
\vspace{10pt}
\caption{Schematic diagram of the system considered}
\label{F:afv:fig1}
\end{figure}

The progress of nanotechnology over the last few years has made it
possible to produce conducting nanostructures in which new physical
phenomena have been observed. Specifically, hybrid structures
consisting of superconductors (S) and normal conductors (N) have
been created using metal films
\cite{afv:r2,afv:r3,afv:r4,afv:r5}
or semiconductor layers
\cite{afv:r1,afv:r6,afv:r7,afv:r8} as the normal conductors. The transport properties
of these S/N structures have turned out to be quite unusual.
First, the subgap conductance (zero-bias anomaly) has been observed in SIN tunnel
junctions
at low temperatures ($T < $100 mK) \cite{afv:r1}(see also \cite{afv:r7,afv:r8}).
Second, conductivity oscillations have
been observed in these
mesoscopic structures in a magnetic field $H$ (i.e., in structures
with dimensions less then the phase-breaking length $L_{\varphi}$).
Oscillations of the conductivity of the N channel appeared if the structure
contained superconducting or normal loops
\cite{afv:r2,afv:r3,afv:r4,afv:r5,afv:r6}.
Moreover, for an N channel in contact with superconductors a nonmonotonic
dependence of the conductivity on the temperature $T$ and voltage
$V$ has been observed at $T{\ll}T_c$
\cite{afv:r5}.
The main experimental facts have been explained in recent theoretical works (see review articles
\cite{afv:r23,afv:r24}).
It was established that the proximity effect plays the main role in the transport
properties. For example, the conductivity of an N channel in
the structure shown in Fig.\ref{F:afv:fig1}
changes as a result of the contribution of the condensate induced by the
proximity effect. Since the condensate is induced by both superconductors in a
nonlocal manner, interference appears and a term $-\delta R\cos \varphi$, which
depends on the phase difference $\varphi$ between the superconductors, arises in
the resistance of the N channel \cite{afv:r9,afv:r10,afv:r11}. The phase difference increases
with the magnetic field $H$, and this results in oscillations of the
conductivity of the N channel in a magnetic field. The nonmonotonic
dependence of the resistance $R$ of an N channel on $T$ and $V$ has
also been explained \cite{afv:r12,afv:r13,afv:r14,afv:r15} (see also the theoretical works in
the Conference Proceedings in Ref.\onlinecite{afv:r14}). The nonmonotonic dependence
of the resistance $R(T,V)$ of a point contact ScN (c is a constriction)
was first obtained theoretically in Ref. \onlinecite{afv:r16}.

New effects have also been predicted in theoretical work
devoted to S/N structures. For example, in Refs.
\onlinecite{afv:r14,afv:r17}
it was shown that the critical Josephson current $I_c$ in a structure
of the type displayed in Fig.\ref{F:afv:fig1}
depends on the voltage $V_S$ between
the S and N conductors, changing sign ( $\pi$ - contact) if $V_S$ exceeds
a certain value. In addition, it has been shown that the Josephson
effect also arises in the case when current flows only through
one S/N boundary. Several different configurations of S/N structures were
studied in Ref.
\onlinecite{afv:r18}, determining that under certain conditions the current-voltage
characteristics of the S/N structures have descending segments ($dI/dV<0$).

An important circumstance was noted in Ref.
\onlinecite{afv:r19}
(see also the works in Ref.
\onlinecite{afv:r14}).
It was shown that the local conductivity of an N channel changes over
distances from the S/N boundary which can be much greater than the coherence
length $\xi_{N}=\sqrt{D/2\pi T}$ ($D$ is the diffusion coefficient).
Important consequences follow from this fact. For example, phase coherence
effects in the conductivity of an N channel remain even if the distance
$2L_1$ between the superconductors is much greater than $\xi_{N}$.
This means that the conductivity oscillations in the structure shown in Fig.\ref{F:afv:fig1}b
will also be observed in the case of a negligibly low critical current $I_c$.
The oscillation conservation effect is due to fact that as $T$ increases,
$I_c$ decreases exponentially
($I_{c}\sim\exp (-2L_{1} /\xi _{N} \left( T\right) $),
and $\delta R$ decreases slowly ($\delta R\sim T^{-1}$) \cite{afv:r20,afv:r35}.

In these lectures we discuss briefly the method of quasiclassical Green's
functions and apply this method to the study of transport phenomena in
mesoscopic S/N structures. We restrict ourselves to the dirty limit where the
mean free path $l$ is essentially less than geometric dimensions of the system
and the coherence length, but exceeds significantly the Fermi wave length
$k_F$ (quasiclassic approximation). We will consider mostly a weak proximity effect,
when the amplitude of the condensate induced in the normal metal is small
compared to the condensate amplitude in the superconductors S. This case occurs
if the S/N interface resistance is larger than the resistance of the normal
conductor N. Results obtained for this case remain qualitatively valid in case when
these resistances are comparable.

In the next Section we present the main necessary equations for the Green's
functions and  a general expression for the current in the N channel in which
a condensate is present due to proximity effect. In Section 3 we will give formulas
describing the subgap conductance of tunnel S/I/N junctions and discuss a possible
physical interpretation of this conductance. In Section 4 we will consider the
conductance of a N channel attached to two superconductors and obtain a formula
describing, in particular, the oscillatory behaviour of the conductance in an
applied magnetic field $H$. Also a nonmonotonic dependence of the conductance
on temperature $T$ and on bias voltage $V$ will be analysed. The possibility of
observing Josephson-like effects in a S/N/S mesoscopic structure (see Fig.\ref{F:afv:fig1}b)
will be considered in Section 5. We will show that zero voltage between
superconductors may exist in some interval of the current through the S/N
interfaces and Shapiro-steps may be observed even in absence of the real
Josephson coupling between superconductors when the distance separating
superconductors exceeds essentially the coherence length
\begin{equation}
 2L_{1}{\gg}\xi_{N}\left(T\right)=\sqrt{D/2\pi T}.\label{E:afv:eq1}
\end{equation}
It is important that these effects arise only in the case
when a current $I$ flows along the N channel and the dissipation takes
place \cite{afv:r21,afv:r35}.\\

\paragraph*{2.Basic equations for quasiclassical Green's functions \protect\\}

The Green's function technique is a powerful tool for the theoretical study of different
phenomena in superconductors and superconducting structures. In the case of superconducting
systems, we need to indroduce condensate Green's functions of the type $<\psi_{\uparrow}\left(1\right)
\psi_{\downarrow}\left(2\right)>$, therefore all the Green's functions have a matrix form. For example,
the retarded (advanced) Green's functions are defined as follows \cite{afv:r22,afv:r23}
\begin{equation}
\hat{G}^{R\left(A\right)}=\pm \theta(t_{1(2)}-t_{2(1)})
\left(\hat{G}^{>}\left(1,2\right)-\hat{G}^{<}\left(1,2\right)\right)
\label{E:afv:eq2}
\end{equation}
Here $\hat{G}^{<}$ and $\hat{G}^{>}$ are
\begin{eqnarray}
\hat{G}^{>}_{\alpha \beta}=-i<\psi_{\alpha}(1) \psi_{\beta}^{+}(2)>(-1)^{\alpha +1},\nonumber\\
\hat{G}^{<}_{\alpha \beta}=i<\psi_{\beta}^{+}(2) \psi_{\alpha}(1)> (-1)^{\alpha +1}
\label{E:afv:eq3}
\end{eqnarray}
We introduced here spin indices in the Nambu-space:
$\psi_{1}\left(1\right)=\psi_{\uparrow}\left(1\right)$,
$\psi_{2}\left(1\right)=\psi_{\downarrow}^{+}\left(1\right)$.  As is well known, the functions
$\hat{G}^{R\left(A\right)}$ describe the excitation spectrum
of the system. In order to describe nonequilibrium processes, one needs
to know the distribution functions which are related to the Green's function
$\hat{G}$ indtroduced by Keldysh. It is convenient to define a supermatrix
$\breve{G}\left(1,2\right)$ elements of which are the matrices
$\hat{G}^{R\left(A\right)}$ = $\left(\breve{G}\left(1,2\right)\right)_{11,(22)}$
and $\hat{G}=\left(\breve{G}\right)_{12}$. The element $\left(\breve{G}\right)_{21}$
is the zero matrix.

In the quasiclassical approximation all components of the Green's functions
$\breve{G}_{ex}\left(1,2\right)$ are integrated over the variable $\xi_{p}=(p-p_{F}) v_{F}$
and in the corresponding equation for $\breve{G}_{ex}\left(1,2\right)$ an expansion
is carried out in the parameters $(p_{F} d)^{-1}$,$(p_{F} l)^{-1}$ or $(p_{F} \xi_{N})^{-1}$,
where $d$ is the thickness of the S or N films, $\xi_{N}$ is the coherence length in the N
conductor and $l$ is the mean free path. The quasiclassical Green's functions are defined by
the relation \cite{afv:r22,afv:r23}
\begin{equation}
\breve{G}\left(\vec{p}/p,\vec{r};t_{1},t_{2}\right)=\left(i/\pi\right)
\int d\xi_{p} \breve{G}_{ex}\left(\vec{p},\vec{r};t_{1},t_{2}\right)
\label{E:afv:eq4}
\end{equation}
The subscript "$ex$" means exact (nonquasiclassical) Green's functions. Therefore the
quasiclassical Green's functions $\breve{G}\left(1,2\right)$ depends on the angle of
momentum on the Fermi surface, on the coordinate $\vec{r}$, and on two times.

In what follows we need an equation for the supermatrix $\breve{G}\left(1,2\right)$
only in the N conductor, having the form
\begin{equation}
 D\nabla\left(\breve{G}\nabla\breve{G}\right)
 +i\epsilon\left[\breve{\sigma}_{z},\breve{G}\right]= 0.
 \label{E:afv:eq5}
\end{equation}
where $\breve{\sigma}_{z}$ is a diagonal supermatrix elements of which are the Pauli matrix
$\hat{\sigma}_{z}$. Eq.(\ref{E:afv:eq5}) may be averaged over the thickness of the N film $d$.
Performing the averaging, we obtain
\begin{equation}
 D\partial_{x}\left(\breve{G}\partial_{x}\breve{G}\right)
 +i\epsilon\left[\breve{\sigma}_{z},\breve{G}\right]=\epsilon_{b}
 \theta (x_{S})\left[\breve{G}_{S},\breve{G}\right].
 \label{E:afv:eq6}
\end{equation}
where the coefficient $\epsilon _{b} $
is a characteristic energy which is proportional to the transmission of the
S/N boundary:
 $\epsilon_{b}=\rho D/2R_{b\Box}d$,
 $R_{b\Box} $
 is the resistance of a unit area of the S/N boundary;
 $\rho $
and
$d$ are the resistivity and thickness of the N film.

When deriving Eq.\ (\ref{E:afv:eq6}), the boundary condition

\begin{equation}
 D\left(\breve{G}\partial_{z}\breve{G}\right)=\left(\epsilon
_{b} d_{N} \right) \left[ \breve{G} ,\breve{G} _{S} \right] \label{E:afv:eq7}
\end{equation}
was used. Here the z-axis is normal to the plane of the S/N interface.
The boundary conditions for the quasiclassical Green's functions
 $\breve{G}$
have been derived in the general case by Zaitsev \cite{afv:r25} and have been
reduced to the simple form (\ref{E:afv:eq7}) by Kupriyanov and Lukichev \cite{afv:r26}
in the dirty case. In the case of a good S/N contact, condition
(\ref{E:afv:eq7}) is reduced to the continuity of the Green's functions at
the S/N interface:
 $\breve{G}=\breve{G}_{S}$.
In the case of a poor contact
 ($\epsilon_{b}\rightarrow 0$),
 condition (\ref{E:afv:eq7})
gives the same result for the current through the S/N interface
as obtained with the aid of the tunneling Hamiltonian
method. However, for a S/N contact with an arbitrary barrier
transparency condition (\ref{E:afv:eq7}) is not applicable. The point is
that when deriving Eq.\ (\ref{E:afv:eq7}) Kupriyanov and Lukichev \cite{afv:r26} restricted
themselves to the Legendre polynomials of the zeroth and first
orders in the expansion of the angle-dependent Green's function
 $\breve{G}$. Meanwhile, one can easily show that all the Legendre harmonics
are excited near the S/N (or $\text{N/N}^{\prime}$) interface. They decay to zero
(except the Legandre polynomials of the zeroth and first order)
over the mean free path away from the interface. In order to
obtain a correction of the next order in
 $\epsilon _{b}$
to condition (\ref{E:afv:eq7}),
one has to solve an integral equation (see Ref.\cite{afv:r27}). In the case of the
S/N interface with an arbitrary barrier transparency, the problem
of boundary conditions for the quasiclassical Green's functions
becomes complicate (boundary conditions and their applicability are discussed in detail
in the Raimondi's Lecture Notes).

Eq.(\ref{E:afv:eq6}) must be solved in the normal conductor  for a particular geometry (see, for example, Fig.\ref{F:afv:fig1})
with boundary conditions at $x=\pm L$. In the case of normal reservoirs the condensate functions
$\hat{F}^{R(A)}\left(\pm L\right)$ are equal to zero and $\hat{G}^{R(A)}\left(\pm L\right)= \pm \hat{\sigma}_{z}$.
In the case of superconducting reservoirs the boundary conditions for the retarded (advanced)
Green's functions are
\begin{equation}
\hat{G}^{R(A)}\left(\pm L\right)=G^{R(A)}\hat{\sigma}_{z} + \hat{F}_{S\pm}^{R(A)}
 \label{E:afv:eq8}
\end{equation}
where $G^{R(A)}=\epsilon / \xi_{\epsilon}^{R(A)}$, $\hat{F}_{S\pm}^{R(A)}=\Delta/\xi_{\epsilon}^{R(A)}\left
(\pm i\hat{\sigma}_{x}sin\phi + i \hat{\sigma}_{y} cos\phi\right), \xi_{\epsilon}^{R(A)}= \sqrt{(\epsilon
 \pm i\Gamma)^{2}-\Delta^{2}}$, $\Gamma$ is a damping in the excitation spectrum in superconductor, $2\phi$
 is the phase difference between superconductors. Eq.(\ref{E:afv:eq8}) is valid if the voltage between superconductors $2V$
 is much less than $\Delta/e$. We assume that there is no barrier between the N conductor and reservoirs.

 The Keldysh function $\hat{G}$ describes the kinetic properties of the system and is related to distribution functions
\begin{equation}
\hat{G}=\hat{G}^{R} \hat{f} - \hat{f} \hat{G}^{A}
 \label{E:afv:eq9}
\end{equation}
where $\hat{f}=f_{o}\hat1+f\hat{\sigma}_{z}$ is a matrix distribution function. The function $f_{o}$ enters an expression for
the supercurrent (in a superconductor it determines the energy gap), and the function $f$ determines the quasiparticle current
(in a superconductor it describes the charge-imbalance and the electric field; see, for example \cite{afv:r28}). In reservoirs
the functions $f_{o}$ and $f$ are supposed to have equilibrium form
\begin{eqnarray}
f_{o}(\pm L)=[tanh((\epsilon+eV)\beta)+tanh((\epsilon-eV)\beta)]/2,\label{E:afv:eq10}\\
f(\pm L)\equiv \pm F_{N}(\epsilon)=\pm [tanh((\epsilon+eV)\beta)-tanh((\epsilon-eV)\beta)]/2,
 \label{E:afv:eq11}
\end{eqnarray}
where $\beta =(2T)^{-1}$.

If the functions $\hat{G}^{R(A)}$ and $\hat{G}$ are known, one can easily find a relation between the applied voltage $2V$ and the
current $I$ in the N conductor. The expression for the current is \cite{afv:r22,afv:r23}
\begin{equation}
I=(\sigma d/8)Tr\hat{\sigma}_{z}\int d\epsilon (\hat{G}^{R}\partial_x\hat{G}+\hat{G}\partial_x\hat{G}^{A})
 \label{E:afv:eq12}
\end{equation}

Eqs.(\ref{E:afv:eq6}) and (\ref{E:afv:eq12}) can be simplified significantly in the case of a weak proximity effect when the amplitudes
of the condensate functions in the N conductor $\hat{F}^{R(A)}$ are small. Then the retarded (advanced) Green's functions
in the N conductor have the form
\begin{equation}
\hat{G}^{R(A)}=G^{R(A)}\hat{\sigma_{z}} + \hat{F}^{R(A)}
 \label{E:afv:eq13}
\end{equation}
where $G^{R(A)} \approx \pm [1+ (F^{R(A)})^{2}/2]$ and $\mid F^{R(A)}\mid\ll 1$. When
obtaining the relation between $\hat{G}^{R(A)}$ and $\hat{F}^{R(A)}$, we employed the normalization condition \cite{afv:r22,afv:r23}
\begin{equation}
({G}^{R(A)})^{2}\hat{1} +(\hat{F}^{R(A)})^{2}=\hat{1},
 \label{E:afv:eq14}
\end{equation}
where
$\hat{1}$ is a unit matrix.

The equation for the condensate functions $\hat{F}^{R(A)}$ follows from Eq.(\ref{E:afv:eq6}) and the expression (14)
\begin{equation}
 \partial _{xx} \hat{F}^{R(A)} - \left(k^{R(A)}\right)^2\hat{F}^{R(A)}=
 -k_{b}^{2} w\left[\hat{F}_{S+}^{R(A)}\delta \left(x-L_{1} \right)
 +\hat{F}_{S-}^{R(A)}\delta \left(x+L_{1} \right)\right]
 \label{E:afv:eq15}
\end{equation}
 where $k_{b}^{2}=2\epsilon_{b}/D,\left(k^{R(A)}\right)^2=(\pm 2i\epsilon + \gamma)/D$, $w$ is the width of the S/N interface
 in the x-direction and is supposed to be much less than $L$ and $\xi_{N}(T)$. We introduce here the depairing rate $\gamma$
 in the N conductor which determines the phase-breaking length $L_{\phi}=\sqrt{D/\gamma}$. Once the functions $\hat{F}^{R(A)}$
 are known from
 a solution of the linear equation (\ref{E:afv:eq15}), we can find the conductance of the N film.

 Let us consider for example the system shown in Fig.\ref{F:afv:fig1}. Writing down Eq.(\ref{E:afv:eq6}) for the matrix element (12), we arrive
 at the equation outside the S/N interface
\begin{equation}
 D\partial_{x}\left[\hat{G}^{R}\partial_{x}\hat{G} + \hat{G}\partial_{x}\hat{G}^{A}\right]
 +i\epsilon\left[\hat{\sigma}_{z},\hat{G}\right]= 0.
 \label{E:afv:eq16}
\end{equation}
Multiplying Eq.(\ref{E:afv:eq16}) by $\hat{\sigma}_{z}$ and calculating the trace, we obtain after the first integration
\begin{equation}
 \left(\partial_{x}f\right)\left[1-m_{-}\right]=J(\epsilon).
 \label{E:afv:eq17}
\end{equation}
Here $J(\epsilon)$ is an $x$-independent constant and
\begin{equation}
 m_{-}=(1/8)Tr\left(\hat{F}^{R}-\hat{F}^{A}\right)^2
 \label{E:afv:eq18}
\end{equation}
is a function which describes the condensate contribution to the N film conductance. The left side of Eq.(\ref{E:afv:eq17})
stems from the first term in the square brackets in Eq.(\ref{E:afv:eq16}) provided that the condensate functions are small.
Therefore, according to Eq.(\ref{E:afv:eq12}), the current $I$ is an integral from the "partial current" $J$
\begin{equation}
 I=(\sigma d/2)\int d\epsilon J(\epsilon)
 \label{E:afv:eq19}
\end{equation}
Solving Eq.(\ref{E:afv:eq17}) with boundary conditions (\ref{E:afv:eq11}), we can find a relationship between the current and voltage $I(V)$.
In the next Sections we will analyse the conductance of S/N mesoscopic systems.
\\

\paragraph*{3.Subgap conductance in SIN junctions \protect\\}

In this Section the subgap conductance in superconductor/insulator/normal metal (S/I/N) tunnel junctions will
be discussed. As is well known from conventional theory for S/I/N junctions, the
subgap conductance should exponentially decrease with decreasing temperature
T (see, for example, Ref. \onlinecite{afv:r29}). However, experiments
on Nb/n+InGaAs contacts have established that a peak in conductance
appears at zero-bias if the temperature becomes low enough ($T\ll\Delta $),
and the magnitude of this peak at low temperatures ($T\approx $ 50 mK)
is comparable with the conductance in the normal state \cite{afv:r1}. This
contact can be considered as a tunnel S/I/N junction. A Schottky
barrier at the interface plays the role of the insulating layer I. An explanation for anomalous
transparency of the SIN junction at low voltages and temperatures
($T,eV\ll\Delta $) was  suggested in Refs. \onlinecite{afv:r9,afv:r30,afv:r31,afv:r32,afv:r33,afv:r34}. According to the interpretation
proposed in Ref. \onlinecite{afv:r32}, the subgap conductance is due to a component
of the current which, in the case of a SIS Josephson junction,
gives the so-called interference current. This component can
be presented in the form
\begin{equation}
 I_{int} =-\left(8R_{b} \right)^{-1} \int d\varepsilon \cdot F_{N} \left(\varepsilon,V\right)
\left(F^{R} +F^{A} \right)\left(F_{S}^{R} +F_{S}^{A} \right),
\label{E:afv:eq20}
\end{equation}
where $F_{N}\left(\varepsilon,V\right)$ is defined in Eq. (\ref{E:afv:eq11});
$F_{S}^{R\left(A\right)} =\Delta /\left(\left(\varepsilon \pm i\Gamma\right)^{2} -\Delta ^{2} \right)^{1/2} $
are the condensate, retarded (advanced) Green's functions in the superconductor. This formula can be obtained
from the general expression for the current (\ref{E:afv:eq12}) and from Eq.(\ref{E:afv:eq6}) in which we have to put $\theta (x_{S})=1$.
If the energy $\varepsilon $ is small, $F^{R} =F^{A} \approx -i$. In the case of a S/I/N junction, $F^{R(A)} $
are the condensate functions in the N electrode. To zero order in
the barrier transmittance (i.e., in $R_{b}^{-1} $), they are equal to zero.
If the proximity effect is taken into account, they differ from
zero and in the case of a planar S/I/N junction they have the form (see, for example, Ref.\onlinecite{afv:r32})
\begin{equation}
F^{R(A)}=\left\{
\begin{array}{cc}
\pm \varepsilon _{b}/\left( \varepsilon \pm i\gamma \right) , & \gamma
> \varepsilon _{b} \\
\varepsilon _{b}/\left[ \left( \varepsilon \pm i\gamma \right)
^2-\varepsilon _{b}^2\right]^{1/2}, & \gamma <\varepsilon _{b}
\end{array}
\right. \label{E:afv:eq21}
\end{equation}
This formula can be obtained from Eq.(\ref{E:afv:eq6}) in the case of weak and
strong proximity effect.
It is seen from Eq.(\ref{E:afv:eq21}) that
 $F^{R(A)} $
 are small if
 $\varepsilon_{b}\ll{\varepsilon ,\gamma }$,
 where
 $\varepsilon $
 is determined either by temperature $T$
 or voltage $V$.
 In the opposite limit when $\varepsilon $ and $\gamma$ are small compared with
 $\varepsilon _{b} $,
 $F^{R(A)} $
 are not small, and the differential conductance normalized by
 $R_{b}^{-1} $,
 $S=R_{b} dI/dV$,
 calculated from Eq.(\ref{E:afv:eq20}) for T=0 and V=0 is not small either.
The integrand in Eq.(\ref{E:afv:eq20}) is not zero if
 $\left| \varepsilon \right| <\Delta $
 because $F^{R} =F^{A} $ at $\left| \varepsilon \right| <\Delta $ and
 $F^{R} =-F^{A} $ at $\left| \varepsilon \right| >\Delta $.
 This means that the current given by Eq.(\ref{E:afv:eq1}) is caused by the
charge-transfer mechanism of the same type as Andreev reflection processes.
The second important circumstance leading to the subgap conductance
is related to an anomalous proximity effect when the amplitude
of the condensate functions
 $F^{R(A)} $
 at small energies
 $\varepsilon $
 is not small.

 The density-of-states (DOS) in the N electrode is changed drastically in the case
 of the strong proximity effect: the DOS is zero at $\mid\epsilon\mid < \epsilon_{b}$ and
 the DOS = $\epsilon/\sqrt{\epsilon^{2}-\epsilon_{b}^{2}}$ in the interval $\Delta \gg \mid\epsilon\mid > \epsilon_{b}$.
  In a one-dimensional S/I/N junction the DOS has a quasigap at
 $\mid\epsilon\mid < \epsilon_{b}$. In both cases of a planar or one-dimensional S/I/N junctions
 the zero-bias, zero-temperature conductance coincides with it's value in the normal state \onlinecite{afv:r32}.
\\

 \paragraph*{4.Conductance of the Andreev interferometer \protect\\}

 Consider the system shown in Fig.\ref{F:afv:fig1} (the Andreev interferometer). In order to calculate the normalized
 differential conductance of the N channel $S=R_{N}dI/vV$ in the presence of a phase difference between
 superconductors, we must solve Eq.(\ref{E:afv:eq17}) taking into account the boundary condition (11). The
 function $m_{-}$ is small by assumption. Therefore the expression for $J$ may be presented in the form
\begin{equation}
 J\left(\epsilon \right)=F_{N} \left(\epsilon ,V\right)\left[1-\left\langle m_{-} \right\rangle\right]/L
\label{E:afv:eq22}
\end{equation}
where
 $\left\langle m_{-} \right\rangle=\left(1/L\right)\int\nolimits_{0}^{L}dx\cdot m_{-}^{2}  $
. Substituting (\ref{E:afv:eq22}) into Eq.(\ref{E:afv:eq19}), we can obtain an expression
for $S$. Here we present the formula for a deviation of the normalized
differential conductance from it's value in the normal state
 $R_{N}$: $\delta S\equiv \left(2L/\sigma d \right)dI/dV-1$
. We obtain
\begin{equation}
 \delta S=-\left(1/2\right)\int d\epsilon \cdot \beta \cdot F_{N}^{'}\left(\epsilon ,V\right)\left\langle m_{-}
\right\rangle,
\label{E:afv:eq23}
\end{equation}
where
 $F_{N}^{'}\left(\epsilon ,V\right)=\left[\cosh ^{-2} \left(\epsilon +eV\right)\beta +\cosh ^{-2} \left(\epsilon
-eV\right)\beta \right]/2$. By virtue of the definition of
 $\left\langle m_{-} \right\rangle$
 (see Eqs.(\ref{E:afv:eq18})), we can write
 $\left\langle m_{-} \right\rangle$ in the form

\begin{equation}
 \left\langle m_{-} \right\rangle =
Tr\left\langle\left(\hat{F}^{R}\right)^2  +
\left(\hat{F}^{A}\right)^2 - 2\hat{F}^{R}\hat{F}^{A}\right\rangle/8
\label{E:afv:eq24}
\end{equation}
The first two terms in (\ref{E:afv:eq24}) determine a change in the
DOS of the N channel due to the condensate (this term reduces
the conductance), and the last, so-called anomalous, term leads
to an increase of the conductance.

\begin{figure}[b!] 
\centerline{\epsfig{file=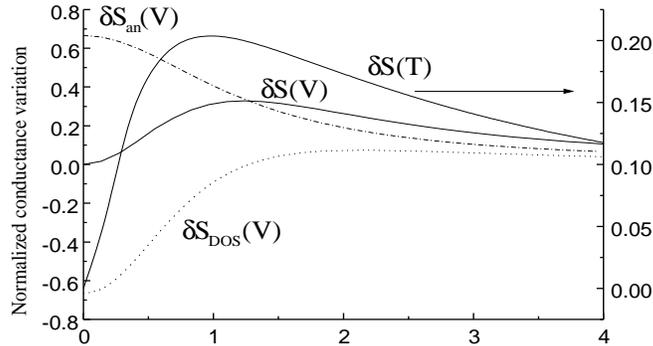,height=1.97in,width=3.5in}}
\vspace{10pt}
 \caption
  {
   Normalized conductance
    $\delta S$
   vs. normalized voltage
    $eV/\epsilon _{L} $
   at
    $T=0$
   and
   vs. normalized temperature
    $T/\epsilon _{L} $
   at
    $V=0$
   for the structure shown in Fig.1a.
  }
 \label{F:afv:fig2}
\end{figure}

As it is seen from Eq.(\ref{E:afv:eq23}), in order to find the conductance,
we need to solve Eq.(\ref{E:afv:eq15}) for the condensate functions $\hat{F}^{R(A)}\left(x\right)$.
In this Section we present here the solution for the geometry shown in Fig.\ref{F:afv:fig1}a.
\begin{equation}
 \hat{F}^{R(A)}\left(x\right) =
 i\hat{\sigma}_{y} F_{S}^{R(A)} r\left[\theta ^{R(A)} \cosh \theta
 ^{R(A)}\right]^{-1} \sinh \left[k^{R(A)} \left(L-\left| x\right| \right)\right]\cdot \cos \varphi
 \label{E:afv:eq25}
\end{equation}
Here
 $r=k_{0}^{2} Lw$ is the ratio of the N channel resistance to the S/N resistance,
 $\theta ^{R(A)} =k^{R(A)} L$. Calculating $\left\langle m_{-} \right\rangle$ we find
\begin{eqnarray}
 \left\langle m_{-} \right\rangle =
 \left(r^{2} /8\right)\left\{ Re \left[\sinh \left(2\theta \right)/2\theta
 -1\right]/\left(\theta \cosh \theta \right)^{2}
 -\left[\sinh \left(2\theta _{1} \right)/2\theta _{1}\right.\right.\nonumber\\
 \left.\left.-\sin \left(2\theta _{2} \right)/2\theta _{2} \right]/\left| \theta \coth \theta \right|
 ^{2} \right\}\left(1+\cos 2 \varphi \right)
 \label{E:afv:eq26}
\end{eqnarray}
where
 $\theta =\theta _{1} +i\theta _{2} $. The first term in (\ref{E:afv:eq26}) determines a contribution to the
conductance due to a change in the DOS, and the second term is
related to the anomalous term ($\hat{F}^{A}\hat{F}^{R} $). In Fig.\ref{F:afv:fig2} we show the dependence
 $\delta S_{DOS} $
(first term contribution) and the
 $\delta S_{an} \left(V\right)$
dependence (anomalous term contribution) at
 $T=0$.
It is seen that
 $\delta S_{DOS} \left(V\right)$
is negative and
 $\delta S_{an} \left(V\right)$
 is positive.
The total change in the conductance
 $\delta S\left(V\right)=\delta S_{DOS} \left(V\right)+\delta S_{an} \left(V\right)$
is shown by the solid line. This quantity increases with increasing
 $V$
from zero, reaches a maximum at
 $V_{m} \approx \epsilon _{L} /e$
, and decays to zero with further increase of
 $V$
(here
 $\epsilon _{L}=D/L^{2} $
 is the so-called Thouless energy).

 As follows from (\ref{E:afv:eq23}) and (\ref{E:afv:eq26}), the conductance
 $\delta S$
oscillates with increasing the phase difference. In Fig.\ref{F:afv:fig2} we also plot
the temperature dependence of the zero-bias conductance. Both curves,
 $\delta S\left(V\right)$
at $T = 0$ and
 $\delta S\left(T\right)$
at $V = 0$ are similar. Note that nonmonotonous temperature dependence of
the conductance was obtained earlier in Ref.\onlinecite{afv:r16} where a short
S$c$N contact was analyzed (here $c$ means a constriction).
\\

\paragraph*{5.Dissipative Josephson-like effects in S/N/S structures \protect\\}

In this Section we discuss a possibility to observe Josephson-like effects in mesoscopic S/N/S structures
(see Fig.\ref{F:afv:fig1}b) with negligible Josephson coupling between superconductors, i.e., when the inequality (1)
is fulfilled \cite{afv:r21,afv:r35}. Following the same steps as in Section 2, we obtain instead of Eq.(\ref{E:afv:eq17})
\begin{equation}
 \left( 1-m_{-} \left( x\right) \right) \partial _{x} f=\left\{
 \begin{array}{cc}
  J+J_{1} -J_{S} , & 0<x<L_{1}  \\
  J, & L_{1} <x<L
 \end{array}
 \right.\label{E:afv:eq27}
\end{equation}
In what follows the function $m_{-}$ plays the most important role.

\begin{figure}[b!] 
\centerline{\epsfig{file=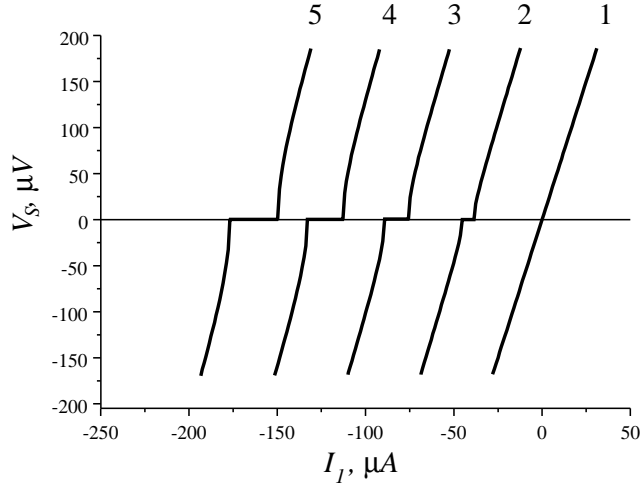,height=2.76in,width=3.5in}}
\vspace{10pt}
\caption{$V_S$ versus the current $I_1$ for the following values of the current:
1 - 0, 2 - 250$\mu\rm A$,
3 - 500$\mu\rm A$, 4 - 750$\mu\rm A$, 5 - 1$m\rm A$. Here
$\delta R_1 =0.1R_1$, $R_b =5R_1$, $R_1 =1 \Omega $.}
\label{F:afv:fig3}
\end{figure}

The current on the segment (0, $L_1$) is determined by Eq.(\ref{E:afv:eq19})
if $J$ is replaced by $J+J_1$. The quantity $J_S$, the superconducting
``current'',  is constant over the segment ($L_1$, $L$) and ($0$, $L_1$) and is equal to
\begin{equation}
 J_{S} =\left( 1/4\right) Tr\hat{\sigma } _{z} \left( \hat{F} ^{R}
 \partial _{x} \hat{F} ^{R} -\hat{F} ^{A} \partial _{x} \hat{F} ^{A}
 \right)\label{E:afv:eq28}
\end{equation}
The integral of $J_S$ (\ref{E:afv:eq28}) over the energy is exponentially small
if the condition (\ref{E:afv:eq1}) is satisfied. As follows from Eq.\ (\ref{E:afv:eq6}), the
constant $J_1$, is related with the Green's function and distribution
function in the superconductor. It can be written in the form
\cite{afv:r11}
\begin{equation}
 J_{1} =J_{q} +\tilde{J} _{S} ,J_{q} =\left( \rho /d_{N} \Re _{b} \right)
 \left[ F_{S} \left( \epsilon \right) -f\left( L_{1} \right) \right]\label{E:afv:eq29}
\end{equation}
Here
$\Re_{b}$ = $R_{b\Box} /w\left[ \nu _{N} \nu _{S}+\left( 1/8\right) Tr\left(
 \hat{F} ^{R}+\hat{F} ^{A} \right) \left(\hat{F}_{S}^{R} + \hat{F}_{S}^{A} \right)
 \right] ^{-1}$
 is the resistance of the S/N boundary per unit length
in the $y$ direction and
 $\nu_{N}$, $\nu_{S}$
- are the density of states in the N and S conductors. It can be shown that for
 $V_{N,S}$
which are small compared with $T/e$, the ``supercurrent'' $\tilde{J}_{S}$,
flowing throw the S/N boundary equals $J_S$. The distribution function $F_S$
is the equilibrium function, i.e., it is identical to the function in
Eq.\ (\ref{E:afv:eq11}), if $V_N$ is replaced by  $V_S$ (we measure voltages from the
point 0, where the voltage is equal to zero). Using the fact that  $m_{-}$ is
small, we can integrate Eq.\ (\ref{E:afv:eq27}) and find the relation of $J$
and $J_q$ with $F_N$ and $F_S$ (see the boundary condition (\ref{E:afv:eq11})).
We obtain the normal currents
\begin{eqnarray}
 \left( d_{N} /\rho \right) J=\frac{\Re _{b} F_{N} +\Re _{1} \left( F_{N}
 -F_{S} \right) }{\Re _{b} \Re +\Re _{1} \Re _{2} },\nonumber\\
 \left( d_{N} /\rho \right) J_{1} \approx J_{q} \left( d_{N} /\rho
 \right) =\frac{\Re _{2} F_{S} +\Re _{1} \left( F_{S} -F_{N} \right) }{\Re
 _{b} \Re +\Re _{1} \Re _{2} }\label{E:afv:eq30}
\end{eqnarray}
Here $\Re_{b}$ is determined in Eq.\ (\ref{E:afv:eq29}); the quantity
 $\Re=\Re_{1}+\Re_{2}$, $\Re_{1,2}=R_{1,2}\left(1+\langle m_{-}\rangle
 \right)$
can be termed the partial resistance. The spatial average
 $\langle m_{-}\rangle_{1,2}$
on the segments (0, $L_1$) and ($L_1$, $L$) gives a decrease in the resistances
on account of proximity effect ($\langle m_{-}\rangle$ is negative). All
resistances in Eq.\ (\ref{E:afv:eq30}) depend on the difference of the phases
 $\varphi$
and on the energy; they can be represented in the form
 $\Re_{b} =R_{b}-\delta\Re_{b}\cos\varphi$
and
 $\Re_{1,2}=R_{1,2}-\delta\Re_{1,2}\cos\varphi$.
The corrections to the resistances
 $\delta \Re _b$
and
 $\delta \Re _{1,2} $
are small in the case of a weak proximity effect. The quantities
$R_b$ and $R_{1,2}$, depend, generally speaking, on the energy
 $\epsilon$ (for example, $\nu_{S}$ depends on $\epsilon $).
We assume, for simplicity, that these quantities do not depend on the energy.
This is valid if it is assumed that the superconductors are gapless
(the results remain qualitatively the same in the case of superconductors
with a gap). Then, integrating Eq.\ (\ref{E:afv:eq30}) over energies, we obtain on the
left-hand side the currents $I$ and $I_1$ (see Eq.\ (\ref{E:afv:eq19})).
Eliminating $V_N$ from the two equations obtained, we find for $V_S$
%
%
%
\begin{eqnarray}
 V_{S}=\hbar\partial_{t}\varphi/2e=\nonumber\\
 I_{1}\left[R_{b}+R_{1}-\left(
 \delta R_{b}+\delta R_{1}\right)\cos 2\varphi\right]
 +I\left(R_{1}-\delta R_{1}\cos 2\varphi\right)
\label{E:afv:eq31}
\end{eqnarray}
Here we employed the Josephson relation; $R_b$ is the resistance
of the S/N boundary, which in the case of zero-gap superconductors
is approximately equal to its value in the normal state. The
resistance $R_1$ is also approximately equal to
 $\rho L_{1} /d_{N}$
(the
 $\varphi $
-independent correction arising from $\langle m_{-}\rangle$  is small and
unimportant). Integrating Eq.\ (\ref{E:afv:eq31}), we obtain a relation
between the average voltage
 $\bar{V}_{S} $
and the fixed currents $I$ and $I_1$.
\begin{equation}
 \bar{V}_{S} =\sqrt{ \left[ \left( I+I_{1} \right) R_{1} +I_{1} R_{b}
 \right] ^{2} -\left[ \left( I+I_{1} \right) \delta R_{1} +I_{1} \delta
 R_{b} \right] ^{2}}\label{E:afv:eq32}
\end{equation}
The function
 $\bar{V} _{S} \left( I_{1} \right)$
 is displayed in Fig.\ref{F:afv:fig3} for different currents $I$.
One can see that for
 $I\neq 0$
this dependence is identical to the current-voltage characteristic of a
standard Josephson contact. In this case the critical current is
\begin{equation}
 I_{c} =I\frac{\delta R_{1} R_{b} -\delta R_{b} R_{1} }{\left( R_{b}
 +R_{1} \right) ^{2} }\label{E:afv:eq33}
\end{equation}
Therefore $I_c$ increases in proportion to the current $I$. We shall
show below that the correction
 $\delta R_{1}$
decreases slowly with increasing temperature ($\delta R_{1}\sim T^{-1}$),
and the correction
 $\delta R_{b} $
is small if the condition
(\ref{E:afv:eq1}) is satisfied. Therefore, for $R_b{\gg}R_1$, we obtain
 $I_{c} \simeq I\delta R_{1} /R_b$.
The maximum current $I$ is limited by the condition that Joule heating be small
and by the condition
 $eV_{N} \simeq eIR{\ll}T$.
In the opposite case
 $\delta R_{1}$
decreases as $V_N$ increases.
If the condition (\ref{E:afv:eq1}) is not satisfied and a finite Josephson
coupling exists between the superconductors, then it is easy
to show that the critical current of the structure equals
 $I_{c}^{*} =\sqrt{I_{c}^{2} +I_{cJ}^{2} }$,
where $I_{cJ}$ is the critical Josephson current. An expression for
 $I_{cJ}$ can be easily obtained with the aid of Eq.\ (\ref{E:afv:eq28}). This
expression is presented in Ref. \onlinecite{afv:r20}. The equilibrium phase
difference
 $\varphi _{0}$
for
 $I_1+IR_1/(R_b+R_1)=0$
equals
 $2\varphi _{0} =-\arcsin (I_{c} /I_{c}^{*}) $.

\begin{figure}[b!] 
\centerline{\epsfig{file=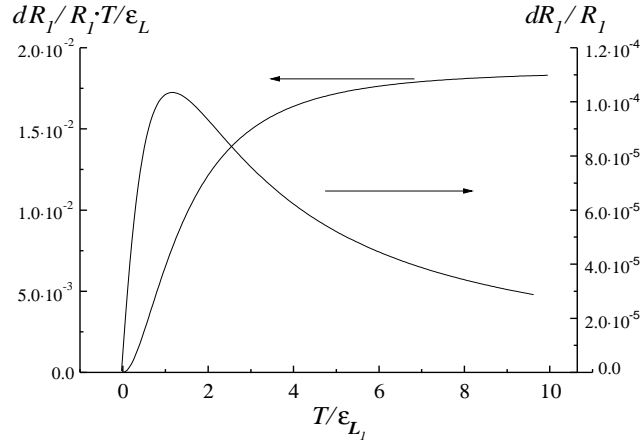,height=2.58in,width=3.5in}}
\vspace{10pt}
\caption{Interference correction $\delta R_1$ to the resistance as a function of
 temperature in the case $L_1 =0.5L$, $R/R_b =0.4$, $\gamma /\epsilon _L=100$,
 $\Delta /\epsilon _L =30$.}
\label{F:afv:fig4}
\end{figure}

To determine
 $\delta R_{1}$
and
 $\delta R_{b}$
it is necessary to find the condensate functions
 $\hat F^{R\left( A\right) } $, For  $|x|<L_1$ the solution of Eq.(\ref{E:afv:eq15}) has
the form
%
%
%
\begin{equation}
 \hat F^{R\left(A\right)}\left(x\right)=F_{_{S}}^{R\left(A\right)}
 \left[i\hat{\sigma}_{y}\cos\left(\varphi \right) P_{y} \cosh\left(kx\right)\right.
 \left.+i\hat{\sigma}_{x}\sin\left(\varphi \right) P_{x} \sinh(kx)\right]^{R\left( A\right)
 }\label{E:afv:eq34}
\end{equation}
Here
 $F_{S}^{R\left( A\right) } $
is the amplitude of the condensate functions in the superconductors. In the
zero-gap case
 $F_{S}^{R\left( A\right) } =\pm \Delta /\left( \epsilon \pm i\gamma _{S}
 \right) $,
where
 $\gamma _{S} $
is the frequency of spin-flip collisions with impurities. The functions
 $P_{x,y}$
equal:
%
%
%
\\
 $P_{x} =b\sinh\theta _{2} /\left( \sinh\theta +b\sinh\theta _{1} \sinh\theta _{2}
 \right)$,\\
 $P_{y} =b\sinh\theta _{2} /\left( \cosh\theta +b\cosh\theta _{1} \sinh\theta
 _{2} \right)$,\\
 $b=\rho w/\left( R_{b{\Box}} d_{N} \right)k$,  $k^{R(A)} =\sqrt{\mp
 2i\epsilon /D}$,
 $\theta =\theta _{1} +\theta _{2}$,
 $\theta_{1,2}\equiv\theta^{\prime}_{1,2} +i\theta^{\prime\prime}_{1,2}
 =kL_{1,2}$
Once the functions
 $\hat{F}^{R\left( A\right) } $
are known, the interference correction
 $\delta R_{1} $
to the resistance can be calculated:
\begin{equation}
 \delta R_{1} =-R_{1} \int\limits_{0}^{\infty }d\epsilon \beta
 \cdot \cosh^{-2} (\epsilon \beta) \langle m_{-}\left(x,\varphi\right)
 -m_{-} \left(x,\pi /2\right) \rangle_{1}\label{E:afv:eq35}
\end{equation}
With the aid of the expressions for $\langle m_{-}\rangle _1$
and for
 $\hat{F}^{R\left( A\right) } $
(see Eq.\ (\ref{E:afv:eq34})),  we find
\begin{equation}
 \delta R_{1} /R_{1} =\int\limits_{0}^{\infty }d\epsilon \beta
 \cdot \cosh^{-2} (\epsilon \beta) M\left( \epsilon \right),
\label{E:afv:eq36}
\end{equation}
where
 $M\left(\epsilon\right)=
 \left(1/8\right)
 \Bigl\{
 \left|F_{S}\right|^{2}
 \left[
 \left|P_{y}\right|^{2}
 \left[
 \sinh\left(2\theta^{\prime}_{1}\right)/2\theta^{\prime}_{1}
 +\sin\left(2\theta^{\prime\prime}_{1}\right)/2\theta^{\prime\prime}_{1}
 \right]
 \right.\Bigr.$
 $\Bigl.\left.-\left|P_{x}\right|^{2}
 \left[\sinh\left(2\theta^{\prime}_{1}\right)/2\theta^{\prime}_{1}
 -\sin\left(2\theta^{\prime\prime}_{1}\right)/2\theta^{\prime
  \prime}_{1}\right]\right]$
  $\Bigl.+\left.Re F_{S}^{2}\left[P_{y}^{2}\left(
  \sinh\left(2\theta_{1}\right)/2\theta_{1}+1\right)\right.\right.\Bigr.$
  $\Bigl.\left.\left.-P_{x}^{2}
  \left(\sinh\left(2\theta_{1}\right)/2\theta_{1}-1\right)\right]
  \right.\Bigr\}$.
The temperature dependence of
 $\delta R_{1} $
is displayed in Fig.\ref{F:afv:fig4}. One can see that for
 $T>\epsilon _{L_{1} } =D/\left( 2L_{1} \right) ^{2} $
the quantity
 $\delta R_{1} $
decreases as
 $T^{-1}$
with increasing temperature. As noted in Refs. \onlinecite{afv:r15,afv:r18},
the slow decrease of
 $\delta R_{1} \left( T\right) $
is due to the so-called anomalous term $F_{R}F_{A}$ in $\langle m_{-}\rangle _1$. The
special role of this term, which is nonanalytic both in the upper
and lower planes of
 $\epsilon $,
was noted in Ref. \onlinecite{afv:r36}.

The Josephson current $I_S$  is determined by the integral of $J_S$
(\ref{E:afv:eq28}), over all energies, i.e., the integral of products of either
advanced or retarded Green's functions. It can be calculated
by closing the integration contour in the upper (lower) half
plane of
 $\epsilon $
and switching to summation over the Matsubara frequencies
 $\omega _{n} =\pi T\left( 2n+1\right) $.
For such energies the functions
  $F^{R(A)}$
decay exponentially over distances
 $k^{-1} \left( \omega _{n} \right) \leq \xi _{n} \left( T\right) $
away from the S/N boundary. Therefore the current $I_S$
will be exponentially small
 ($I_{S}\sim\exp \left( -2L_{1} /\xi _{N} \left( T\right) \right) $.
The function $I_{S}(T)$ for the structure shown in Fig.\ref{F:afv:fig1}b is presented
in Ref. \onlinecite{afv:r20}. Similar arguments are also applicable to the
calculation of
 $\delta R_{b} $,
since for
 $T<\gamma _{S} $
the functions
 $F_{S}^{R} $
and
 $F_{S}^{A} $
can be assumed to be equal and independent of the energy. At the same time,
the function $F^{R}F^{A}$, appearing in the expression for
 $\delta R_{1} $,
decreases over a small (compared with $T$) energy
 $\epsilon _{L_{1} } =D/\left( 2L_{1} \right) ^{2} $
and makes a nonzero contribution. For such energies
the characteristic decay length of
 $F^{R(A)}(x)$
is of the order $L_1$, i.e., of the order of the distance between the
superconductors.

In order to observe long-range Josephson effects, the critical
current $I_c$ must exceed the fluctuation current
 $Te/\hbar $:
 $I_{c} {\gg}Te/\hbar $. On the other
hand, the ordinary Josephson effect is negligible if the condition
 $\epsilon _{L_{1} } {\ll}T$
is fulfilled. Combining these inequalities, we obtain the condition
\begin{equation}
 TR_{b} R_{1} /\left( \delta R_{b} R_{Q} \right){\ll}\epsilon _{L_1}
 {\ll}T\label{E:afv:eq37}
\end{equation}
which should be satisfied for observation of the effects under consideration.
Here
 $R_{Q} =\hbar /e^{2} \approx 3k\Omega $,
and we took into account that a maximal value of $I$ is determined by the
relation
 $eIR\leq \epsilon _{L_{1} } $.
Otherwise
 $\delta R_{1} $
decreases with increasing $I$. The first inequality of (\ref{E:afv:eq37}) means that
the zeroth Shapiro step on the
 $I_{1} \left( V_{S} \right) $
curve is absent at $I=0$. If the second inequality of (\ref{E:afv:eq37}) is not
fulfilled, then the critical current is not zero at
 $I=0$
(ordinary Josephson effect). In this case the effective critical current
 $I_{c}^{*} $
should first increase with increasing $I$ and then decrease when $I$ exceeds
 $\epsilon _{L_{1} } /eR$.\\

\paragraph*{6.Conclusion\protect\\}

In conclusion we note  that, as one can see from Fig.\ref{F:afv:fig4}, the
correction
 $\delta R_{1} $
to the resistance of the normal channel caused by the proximity effect depends on the
temperature $T$ in a nonmonotonic way: it is equal to zero at $T=0$ (the bias
voltage is zero as well), reaches a maximum at
 $T\approx \epsilon _{L_{1} } $
and decays to zero at higher $T$. Such behavior of
 $\delta R_{1} \left( T\right) $
is related, as noted in \cite{afv:r15}, to different dependencies of two
contributions to
 $\delta R_{1} $
on the energy
 $\epsilon $. One contribution which
increases the N channel resistance is connected with a decrease of the
density-of-states in the normal channel, which is described by the last term in
 $M\left( \epsilon \right) $
(see Eq.\ (\ref{E:afv:eq36})). Another contribution (anomalous) which diminishes
the resistance of the normal channel is described by the first two terms in
$M\left( \epsilon \right) $. This contribution exactly compensates
a contribution due to a change in the density-of-states of the normal channel at
 $\epsilon =0$
and dominates at
 $\epsilon \neq 0$.
At $T>T_{c} $ it leads to the Maki-Thompson contribution to the
paraconductivity. Mathematically, compensation of the two contributions at
 $\epsilon =0$ arises because at $\epsilon =0$ $F^{R} =F^{A} $ and $m_{-} $
in Eq.\ (\ref{E:afv:eq35}) tends to zero. The nonmonotonic behavior of $\delta R$
has been observed in an experiment \cite{afv:r5}. It would be interesting to observe
the long-range Josephson effect experimentally.

This work was supported by the Russian Fund for Fundamental Research
(Project 96-02-16663a), by the Russian Grant on high $T_c$ Superconductivity
(Project 96053), and by CRDF Project (RP1-165). This support is gratefully
acknowledged.

%
%
%
%
\end{document}